\documentclass[aps,prb,twocolumn,floatfix]{revtex4-1}

\usepackage[utf8]{inputenc}
\usepackage{amsfonts}
\usepackage{amsmath}
\usepackage{amssymb}
\usepackage{bbm}
\usepackage{amsthm}
\usepackage{empheq}
\usepackage{enumitem}
\usepackage{verbatim}
\usepackage{graphicx}
\usepackage{epstopdf}

\usepackage{subfigure}
\usepackage{hyperref}
\usepackage{placeins}

\begin{document}

\title{Active learning phase boundaries of a quantum many-body system}
\author{Steve Keeling$^{1}$}
\affiliation{$^1$Department of Physics and Astronomy, College of Sciences,\\George Mason University, Fairfax, VA 22030, USA}
\date{\today}


\begin{abstract}
We describe how to use techniques from the field of Machine Learning to direct a variational energy minimization scheme to search for phase boundaries of a quantum many-body system.  The modeled physical system presents states of finite momentum condensate (FMC) first described by Fulde and Ferrell\cite{fulde1964superconductivity} and separately by Larkin and Ovchinnikov\cite{larkin1965inhomogeneous} (also known as FFLO/LOFF states), as well as a uniform superfluid phase\--all of which is interesting in its own right; however, a full description of the multitude of phase boundaries is expensive from a computational standpoint.  We treat the output of the energy minimization scheme as a labeled synthetic data set to train a support vector classifier (SVC) to separate states of FMC from superfluid and normal states. We can then use the trained SVC to refocus the minimizer to intensify its calculations near the boundary separating each of the three regions.  Doing so will preclude using the minimizer to perform expensive calculations deep within normal or superfluid regions, resulting in a more efficient use of compute time.  The application of the procedure we describe is straightforward and should be applicable in any computational search of distinct phase boundaries.
\end{abstract}

\maketitle

\section{Introduction} \label{intro}
Techniques from the field of machine learning (ML) have proven to be successful in areas as diverse as accuracy in classifying hand-drawn digits\cite{lecun1998gradient} and recognizing facial expressions\cite{dailey2002empath}.  The key concept tying these unrelated areas together is pattern recognition.  That is, analyzing data for relationships that can provide insight into its underlying structure.  Analyzing data for patterns is not new to physics, beginning in earnest in the 16\textsuperscript{th} century with Copernicus, Brahe, Kepler and Galileo.  However, the speed and memory capacity of the modern digital computer has provided physics the ability to automate to some extent the discovery of underlying relationships hidden within data.  Indeed algorithms and techniques from the ML field are being actively applied to quantum many-body physics\cite{carleo2017solving,zhang2018machine,schuld2019quantum,jinnouchi2019phase}.

Our aim here is a hybrid approach that uses an ML algorithm known as a support vector classifier to direct a minimizer to search its parameter space for phase boundaries.  There is recent work in Active Learning (AL) using unsupervised algorithms which work on unlabeled data.  In these cases, neural networks can be shown to fit a non-linear multi-dimensional function\cite{yao2019active}, as well as for efficient quantum information retrieval\cite{ding2019retrieving}.  However, we focus here on supervised learning which requires a labeled data set. In what follows, we briefly discuss the physical model and the variational energy minimization scheme used to perform calculations to find low energy states.  We provide an overview of support vector machines that is tailored to our physical model and energy minimization scheme, and discuss how ML can be used to find parameter values near and within the FMC to make efficient use of compute time.

\section{Phase Boundary Classification}
\subsection{Model Hamiltonian}
The physical system being modeled is a gas of ultracold fermions on a two-dimensional square lattice in the presence of a Zeeman field. We consider an attractive Hubbard Hamiltonian restricted to nearest-neighbor hopping.  In grand canonical formalism, our model Hamiltonian is
\begin{align}
   \hat{\mathcal{H}} = &\sum_{\mathbf{k}\sigma}{ \xi_{\mathbf{k}}c_{\mathbf{k}\sigma}^{\dagger}c_{\mathbf{k}\sigma}}
      - h \sum_{  \mathbf{k}}{ \left( c_{\mathbf{k}\uparrow}^{\dagger}c_{\mathbf{k}\uparrow} -
         c_{\mathbf{k}\downarrow}^{\dagger}c_{\mathbf{k}\downarrow} \right)}  + \nonumber \\
	      &+ \sum_{  \mathbf{k}}{ \left( \Delta_{\mathbf{q}}^* c_{\mathbf{q} - 
                    \mathbf{k} \downarrow} c_{\mathbf{k} \uparrow} + \Delta_{\mathbf{q}} c_{- \mathbf{q} +
		      \mathbf{k} \downarrow}^{\dagger} c_{\mathbf{k} \uparrow}^{\dagger} \right) } \nonumber \\
                        &\qquad + \sum_{\mathbf{k}}{\frac{\Delta_\mathbf{k}^*\Delta_\mathbf{k}}{U}}; \label{modelHamiltonian}
\end{align}
where all sums are restricted to wavevectors $\mathbf{k}$ in the first Brillouin zone; $\xi_{\mathbf{k}} = \epsilon_{\mathbf{k}} - \mu$ with dispersion $\epsilon_{\mathbf{k}} = - 4 t \left( \cos k_x a + \cos k_y a\right)$, with $\mu$ the chemical potential and $h$ the Zeeman field strength. The operators $c_{\mathbf{k} \sigma}^{\dagger}, c_{\mathbf{k} \sigma}$ create or respectively annihilate a fermion with wavevector $\mathbf{k}$ and (pseudo)spin $\sigma \in {\uparrow, \downarrow}$; the order parameter $\Delta_{\mathbf{q}}$ is a $c-$number which depends on momentum; and finally, $U$ parameterizes the on-site interaction strength. We set the lattice spacing $a$ and hopping $t$ to unity and work in units where $\hbar = 1$.

\subsection{Minimization Procedure}
Energy minimization is performed at zero temperature with the chemical potential ($\mu$), Zeeman field ($h$), and interaction strength ($U$) as input parameters. The codes use the given parameters to search phase space for an order parameter of the form:
\begin{equation}\label{eq:OP}
    \Delta({\bf Q}) = \sum_{n=0}^{q-1} \Delta_n e^{i n {\bf Qr}}
\end{equation}
which minimizes the energy; where $\Delta_n$ are complex pairing amplitudes living at integer multiples of the ordering wavevector $\mathbf{Q}$ in the first Brillouin zone. If only $\Delta_0$ is non-zero, the finite-momentum condensate is a plane wave state (PWS) that breaks the time-reversal and rotational symmetries, but not translational symmetry (all local observables are invariant under translations, although the symmetry-broken U(1) phase is not). The latter allows an easy embedding of the PWS order parameter $\Delta({\bf r}) = \Delta_0 e^{i{\bf Qr}}$ with an arbitrary incommensurate $\bf Q$ in the Bogoliubov de-Gennes (BdG) Hamiltonian:
\begin{equation}\label{eq:HBdGPWS}
H_{\textrm{BdG}}^{\textrm{(PWS)}} = \left( \begin{array}{cc}
  \epsilon_{\bf k} & \Delta_0 \\
  \Delta_0^* & -\epsilon_{{\bf Q}-{\bf k}}
\end{array} \right) + h\times\mathbbm{1} \ ;
\end{equation}
where
\begin{equation}
\epsilon_{\bf k} = 2t\sum_{i=x,y} \Bigl\lbrack 1-\cos(k_i) \Bigr\rbrack - \mu
\end{equation}
is the energy of a free fermion with momentum ${\bf k} = (k_x, k_y)$ in the units where the lattice constant is $a=1$.

If multiple amplitudes $\Delta_n$ are non-zero, then the order parameter (\ref{eq:OP}) produces a pair density wave (PDW) that breaks translation symmetry. Incommensurate PDWs are not dynamically stable, so we restrict the ordering wavevector to commensurate values such that:
\begin{equation}
    q\mathbf{Q} \in \mathbb{G}
\end{equation}
for a positive integer $q$, where $\mathbb{G}$ is the set of all reciprocal lattice vectors. The smallest possible value of $q$ is the number of sites in the enlarged unit-cell of the periodic order parameter. The representation of the BdG Hamiltonian is also enlarged $q$ times:
\begin{widetext}
\begin{equation}\label{eq:HBdGPDW}
H_{\textrm{BdG}}^{\textrm{(PDW)}} = \left( \begin{array}{ccccccccc}
  \epsilon_{\bf k} & \Delta_0   & 0 & \Delta_{q-1}   & 0 & \Delta_{q-2}   & \cdots   & 0 & \Delta_1  \\
  \Delta_0^* & -\epsilon_{-{\bf k}}  & \Delta_1^* & 0   & \Delta_2^* & 0   & \cdots   & \Delta_{q-1}^* & 0  \\
  0 & \Delta_1   & \epsilon_{{\bf Q}+{\bf k}} & \Delta_0   & 0 & \Delta_{q-1}    & \cdots   & 0 & \Delta_2  \\
  \Delta_{q-1}^* & 0   & \Delta_0^* & -\epsilon_{(q-1){\bf Q}-{\bf k}}  & \Delta_1^* & 0    & \cdots   & \Delta_{q-2}^* & 0  \\
  0 & \Delta_2   & 0 & \Delta_1   & \epsilon_{2{\bf Q}+{\bf k}} & \Delta_0   & \cdots   & 0 & \Delta_3  \\
  \Delta_{q-2}^* & 0   & \Delta_{q-1}^* & 0   & \Delta_0^* & -\epsilon_{(q-2){\bf Q}-{\bf k}}  & \cdots   & \Delta_{q-3}^* & 0  \\
  \vdots & \vdots & \vdots & \vdots & \vdots & \vdots   & \ddots   & \vdots & \vdots \\
  0 & \Delta_{q-1}   & 0 & \Delta_{q-2}   & 0 & \Delta_{q-3}   & \cdots & \epsilon_{(q-1){\bf Q}+{\bf k}} & \Delta_0 \\
  \Delta_1^* & 0   & \Delta_2^* & 0   & \Delta_3^* & 0   & \cdots & \Delta_0^* & -\epsilon_{{\bf Q}-{\bf k}}
\end{array} \right) + h\times\mathbbm{1} \ ,
\end{equation}
\end{widetext}

After running the minimization procedure, points where the free energy is zero are labeled magnetized metal (MM) states.  Points where the free energy is negative and $\Delta_{0}$ is non-zero are labeled uniform superfluid (USF).  Although above we make the distinction between plane wave states and pair density wave states, points where the free energy is negative, one or more $\Delta_{n}$ are non-zero and either or both components of the ordering wavevector $\mathbf{Q}$ are non-zero are labeled FMC.

\subsection{Synthetic Data}
The synthetic data set we use to train the SVC was originally collected to locate states of FMC and classify not just the USF/FMC and FMC/MM boundaries, but also boundaries between physically different FFLO states.  The calculations required several months of compute time, were performed in stages, and the ranges for the parameters for the minimization scheme reflect the attempt to manually reduce the number of computations.  The ranges for the interaction strength, chemical potential and Zeeman field are:
\begin{itemize}
    \item $U \in [0.1,0.15,0.2,0.25,0.5]$
    \item $\mu \in [0,3.75]$ with a step of 0.25
    \item $h \in [0,3.2]$ with a step of 0.01
\end{itemize}
The inverse of the interaction strength gives the Cooper pairing strength with this range of $U$, providing a compromise between an exceedingly strong pairing that is certain to exhibit FFLO states and a pairing so weak that Cooper pairs will not form in even the weakest of Zeeman fields.  The chosen range for the chemical potential takes advantage of the model's particle-hole symmetry where the phase diagram for each interaction strength is symmetric about half-filling at $\mu = 4.0$, which prevents us from having to perform calculations over the range $\mu \in [0,8]$ where we expect to find FMC states.  The fine mesh in the Zeeman field was used to best highlight the multitude of FFLO phases within the FMC. 
\begin{table}
    \centering
    \caption{Number of calculated/sampled points for each state and each interaction strength.  FMC states make up roughly 10\% of all states found, both before and after sampling.}
    \begin{tabular}{|c|c|}
        \hline
        State (Class) &  No. of Points (Full/Sampled) \\
        \hline
        FMC & 2000/129 \\
        \hline
        MM & 8756/554 \\
        \hline
        USF & 8798/629 \\
        \hline
    \end{tabular}
    \label{table:sample_points}
\end{table}

We imagine using an SVC early in the search for phase boundaries and simulate having fewer calculation points by sampling over the Zeeman field. Taking every tenth point effectively reduces the step size in the Zeeman field from 0.01 to 0.1. Table \ref{table:sample_points} shows how sampling affects the number of points in each state. The FMC states we seek make up just over 10\% of the full data set while, after sampling, FMC states make up just under 10\% of the total.  Sampling also has the effect of reducing the contribution of MM states by 2.5 percentage points at the cost of increasing the contribution from USF points by roughly the same amount. Even though sampling has removed a large number of points, we see the stratified nature of the state labels remains intact. 

\subsection{Support Vector Machines}
There are many references\cite{hastie2003elements,vapnik2013nature,bishop2016pattern} and tutorials\cite{MITOpenCourseWare} on the formulation of support vector machines (SVM); so, this section will serve to highlight the formulation for our particular problem. An SVM\cite{cortes1995support} is a supervised machine learning algorithm devised to find the hyper-plane that separates the positive samples (FMC) from the negative samples (USF/MM).  Supervised is used to mean the data points making up the set must each be labeled prior to training the SVM.  Because the SVM can be formulated as a convex optimization problem, the hyper-plane that is found is a global solution.  In the ML literature, the hyper-plane is referred to as the decision boundary or surface, the positive and negative sample points closest to the decision boundary are called support vectors, and the perpendicular distance from a support vector to the decision surface is called the margin. If we denote the width of the margin $\mathbf{w}$, then the SVM solves the quadratic programming problem
\begin{equation*}
    \arg \min_{\mathbf{w}} \frac{1}{2} ||\mathbf{w}||^2
\end{equation*}
by introducing the Lagrange multipliers
\begin{equation*}
    L\left( \mathbf{w},b,\mathbf{\lambda} \right) = \frac{1}{2} ||\mathbf{w}||^2 - \sum_{i=1}^{N}{\lambda_{i}\left[ y_{i} \left( \mathbf{w} \cdot \phi(\mathbf{x}_{i}) + b  \right) - 1 \right]}
\end{equation*}
with constraints given by
\begin{align*}
    \mathbf{w} &= \sum_{i=1}^{N}{\lambda_{i}y_{i} \phi(\mathbf{x}_{i}})  \\
    0 &= \sum_{i=1}^{N}{\lambda_{i}y_{i}}
\end{align*}
The total number of triples $\mathbf{x}_{i} = (U,\mu,h)_{i}$ comprising the data set is $N$, $b$ is a bias term, $\phi$ is a point transformation, and $y_{i} \in \{0,1\}$ for \textit{not in class} and \textit{in class}, respectively.

In general, phase boundaries of a quantum many-body system will not be linearly separable.  However, we can use what is known as the kernel trick and take advantage of the fact that the minimization problem depends upon the dot product through the distance by making a transformation $\mathbf{x}_{i} \rightarrow \phi(\mathbf{x}_{i})$.  We can also relax the constraints somewhat by allowing some points to lie within the margin or even on the wrong side of the decision surface by introducing slack variables $\xi_{i}$, and minimizing
\begin{equation}
    C \sum_{i = 1}^{N}{\xi_{i}} + \frac{1}{2} ||\mathbf{w}||^2
\end{equation}
where $C$ controls the trade-off between the slack-variable penalty and the margin\cite{bishop2016pattern}. 

\FloatBarrier
\section{Results}
\begin{figure*}[hbt!]
\includegraphics[height=1.5in]{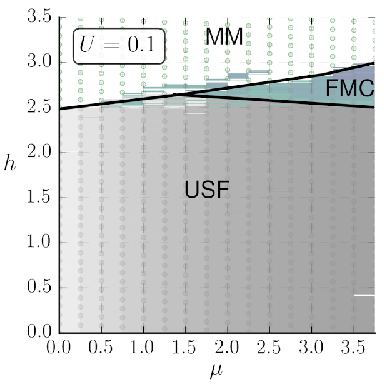}
\includegraphics[height=1.5in]{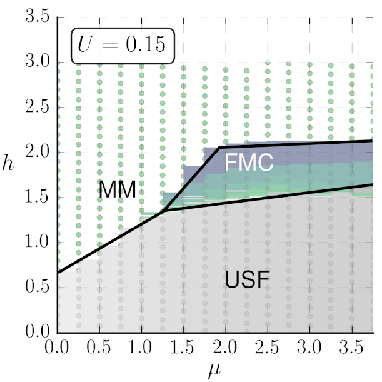}
\includegraphics[height=1.5in]{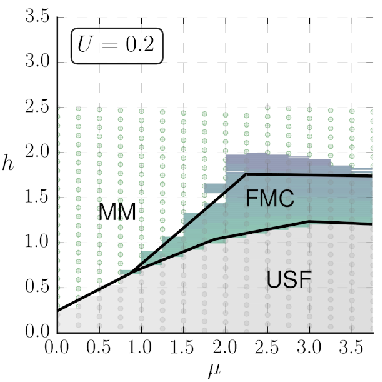} 
\includegraphics[height=1.5in]{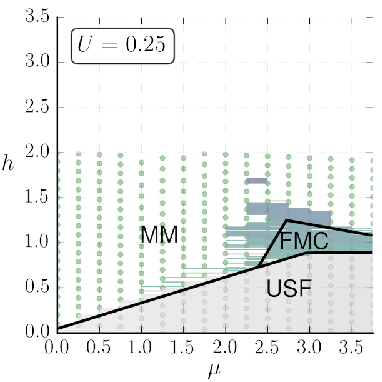} \hskip 0.5in
\includegraphics[height=1.5in]{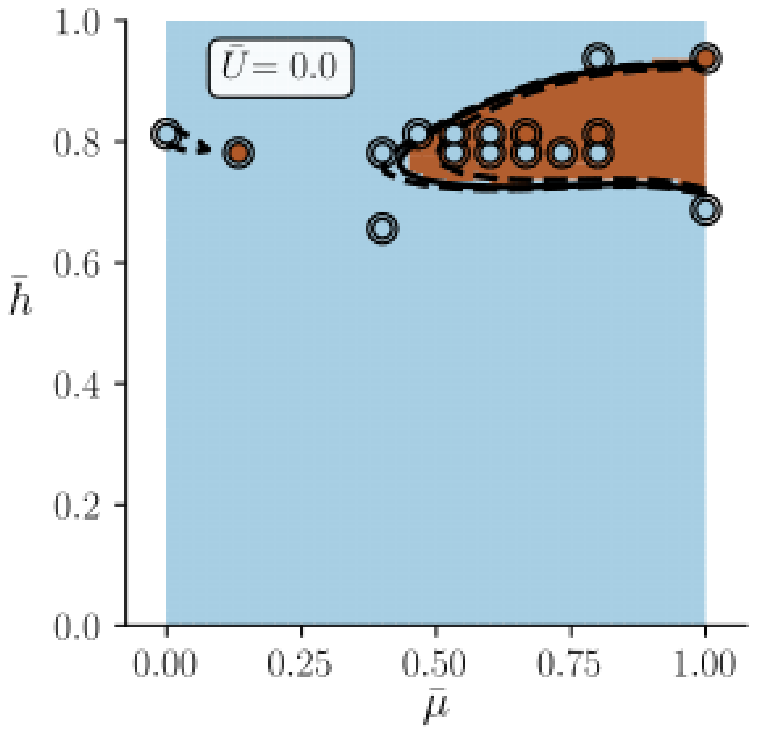} 
\includegraphics[height=1.5in]{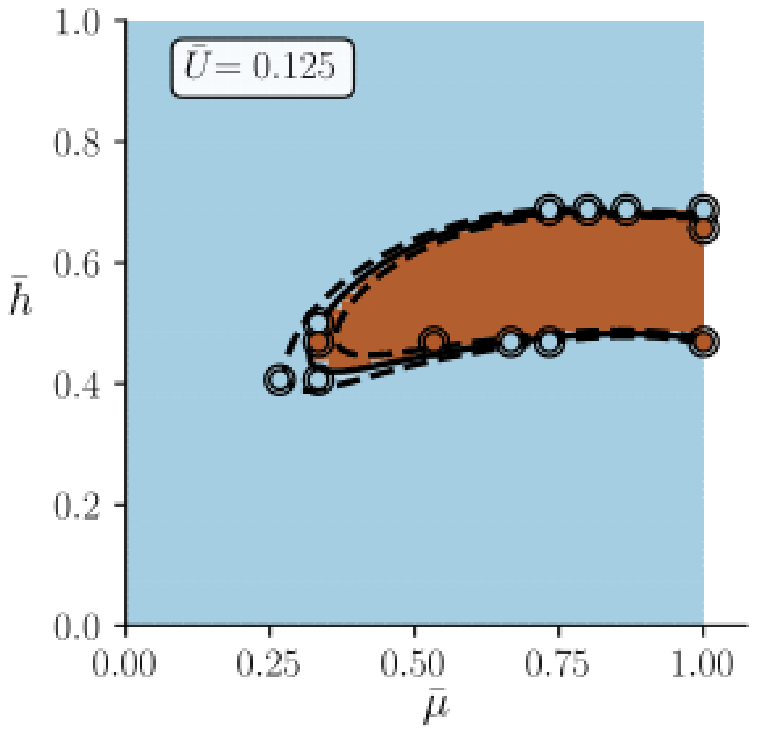} 
\includegraphics[height=1.5in]{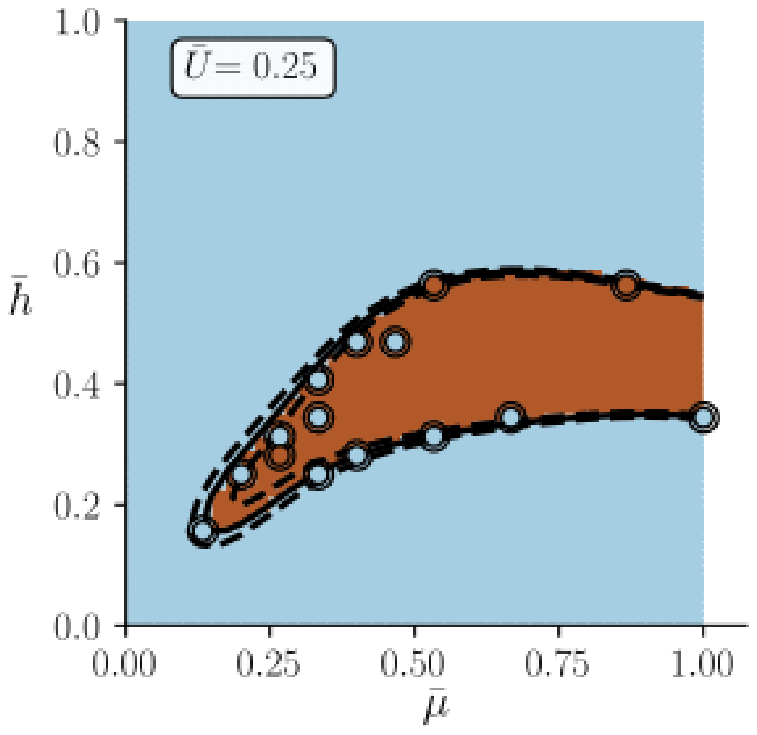} 
\includegraphics[height=1.5in]{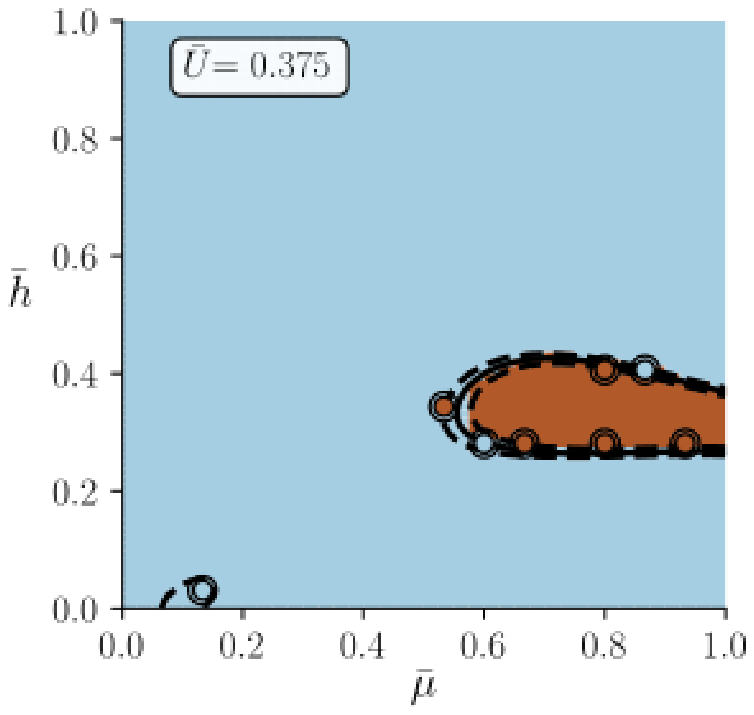} 
\caption{\label{fig:svc-fmc} The top panel shows the the full data set with black hand-drawn boundaries separating USF, FMC and MM states, and green dots representing the sampled points. The grey and blue-green gradients represent the magnitudes of the order parameter and ordering wavevector, respectively, but aren't important for locating the boundaries. Axes are in the parameter ranges for $\mu$ and $h$, and the interaction parameter is shown in the box at the top-left of each plot. The bottom panel is the decision surface and region of FMC (brown). The solid line is the decision surface separating FMC and non-FMC points, and the dashed lines are the margin; the dots are the support vectors with fill color to match the margin they support. Axes are in the normalized ranges for the parameters $\mu$ and $h$, and the box at the top-left of each plot shows the value of the normalized interaction strength.}
\end{figure*}

Our main result is a decision surface for the SVC that approaches the FMC boundary present in the mean-field phase diagram shown in figure (\ref{fig:svc-fmc}).    The top panel shows the entire data at four values of the interaction. The physically interesting FMC states are shown on the top panel in the blue-green regions with USF states in gray, and MM states uncolored.  The bottom panel shows the decision surface for the region of FMC found by the SVC for the respective interaction strength on the top panel. What is most striking about these diagrams is how the SVC is able to create a decision surface that roughly coincides with the phase boundaries using only 7\% of the entire data set. We do not have detailed run times for the for the full data set; however, the SVC has used significantly fewer points to construct a decision surface. This directly translates to a significant decrease in the points to give to the minimization procedure.

\subsection{Active learning process}
Active learning provides us the ability to automate the process of reviewing output and then modifying the input parameters based on that review to refine a computation.  In the particular case of using a minimization procedure to find phase boundaries, the trained and tested SVC provides two important bits of information.  First, it provides a decision surface which can be associated with the phase boundary of the physical system. Second, it provides all of the points that have been incorrectly classified relative to the decision surface.  The most difficult points for an SVC to correctly classify are those near a boundary; so, the incorrectly classified points and those points nearby are obvious choices to include in a new calculation.   
\begin{figure}[hbt!]
\includegraphics[height=2.5in]{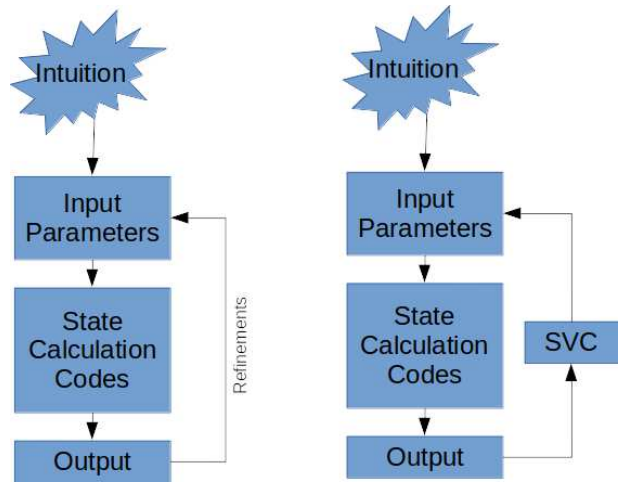}
\caption{\label{fig:alp} The diagram on the left is a cartoon depiction of the creation of the full data set.  Physical intuition is used to define the initial ranges for the $(U,\mu,h)_{i}$. The minimization procedure is executed using these ranges, the output is reviewed to determine whether FMC is found, and then new ranges or steps in those ranges are defined. The minimization procedure is then rerun.  This process is repeated until a satisfactory account of the physical phenomena is made.  The active learning cartoon on the right shows the process to be the same until refinements have to be made manually based on a review of the output. }
\end{figure}

Figure (\ref{fig:alp}) is a cartoon of how an SVC modifies the current analysis. Rather than performing refinements to parameter ranges manaully, an SVC can be used to determine where best to search for FMC.  Though not implemented for this study, a more sophisticated approach would also encode how to modify the tolerance and error thresholds as well as other parameters specific to the minimization procedure to improve confidence in the calculation.

\subsection{Performance}
We use the scikit-learn\cite{pedregosa2011scikit} Python API\cite{buitinck2013ecml} to process the synthetic data by first normalizing the features so that each has a range between 0 and 1.  We also one-hot-encode the target class and make the transformations
\begin{itemize}
    \item $\text{USF} \rightarrow [1,0,0]$
    \item $\text{MM} \rightarrow [0,1,0]$
    \item $\text{FMC} \rightarrow [0,0,1]$
\end{itemize}
The strategy is to use a one-vs-rest classifier that uses a support vector classifier as an estimator in order to perform multi-class classification.  This will produce three binary classifiers for each of the three classes that classify data points as lying inside or outside their class.  For example, data points for the FMC classifier are treated as being labeled either FMC or not-FMC, while data points for the USF classifier are treated as being labeled USF or not-USF, and similarly for MM states. After the synthetic data is processed, we tune the hyper-parameters of the support vector classifier by implementing a randomized cross validated search.  We choose a polynomial kernel for the support vector classifier defined to be
\begin{equation}
    \left( \gamma \langle \phi(\mathbf{x}'),\phi(\mathbf{x}) \rangle + r \right)^d
\end{equation}
with $\gamma$ chosen from a uniform distribution between 0.1 and 10, $r$ chosen from a uniform distribution between 0 and 30, $d \in \left\{3,4,5\right\}$ and the regularization parameter $C$ is chosen from a uniform distribution between 0 and 500.  For cross validation, we specified 30 folds of a stratified shuffle split, and hold out $20\%$ of the data for testing at each fold.  Scoring for the 10 sets of hyper-parameters chosen by the randomized cross validation algorithm were scored for true negative, false positive, false negative, false positive and macro-F1 and a refit of the model was performed using the macro-F1 score.  The macro-F1 is calculated by finding the precision and recall for each class, then finding the unweighted mean of the metrics. Based on the macro-F1 score, the SVC with the best performance uses a polynomial kernel of degree 5, with a $\gamma$ of 5.9526, an $r$ of 18.2351 and a regularization parameter $C$ of 482.5143.

\begin{figure}[hbt!]
\includegraphics[height=1.5in]{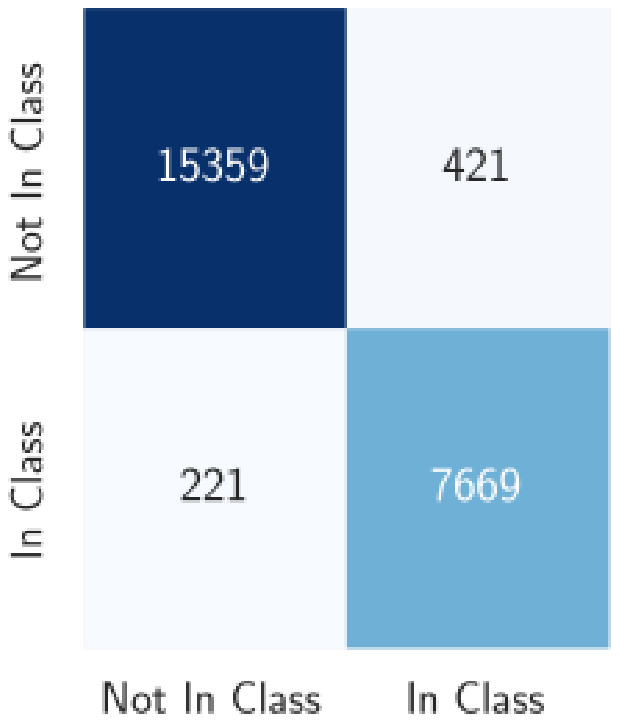} \hskip 0.1in
\includegraphics[height=1.5in]{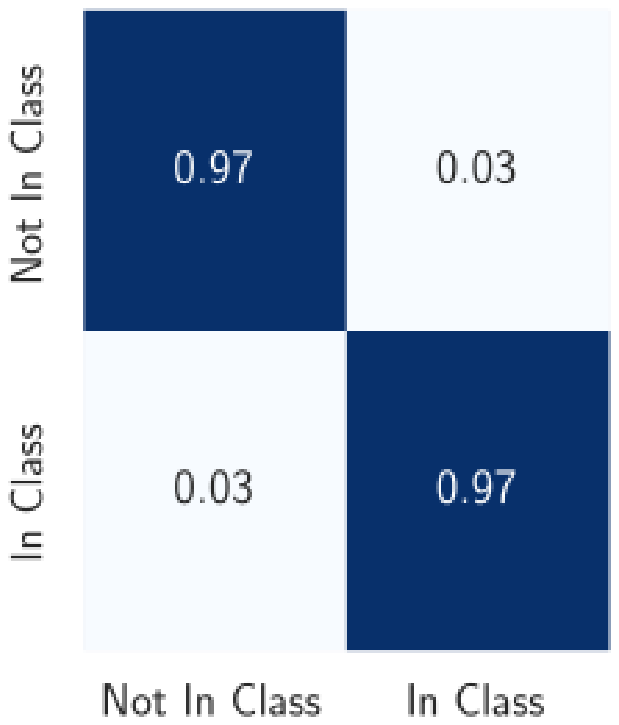}
\caption{\label{fig:best-cm} Confusion matrix for the best performing set of hyper-parameters for the SVC based on the macro-F1 score. The matrix on the left gives the sum over all test runs of true negatives (top left), false positives (top right), false negatives (bottom left), and true positives (bottom right).  The matrix on the right shows the fractions for the raw values. In this case, the top row of the matrix shows the SVC correctly classified 97\% of points outside the class as being outside the class while incorrectly classifying 3\% of the points lying outside the class as belonging to the class. On the bottom row, the SVC incorrectly classified 3\% of the class points as lying outside the class while 97\% of points in the class were correctly classified as such.}
\end{figure}

\begin{figure}[hbt!]
\includegraphics[height=1.5in]{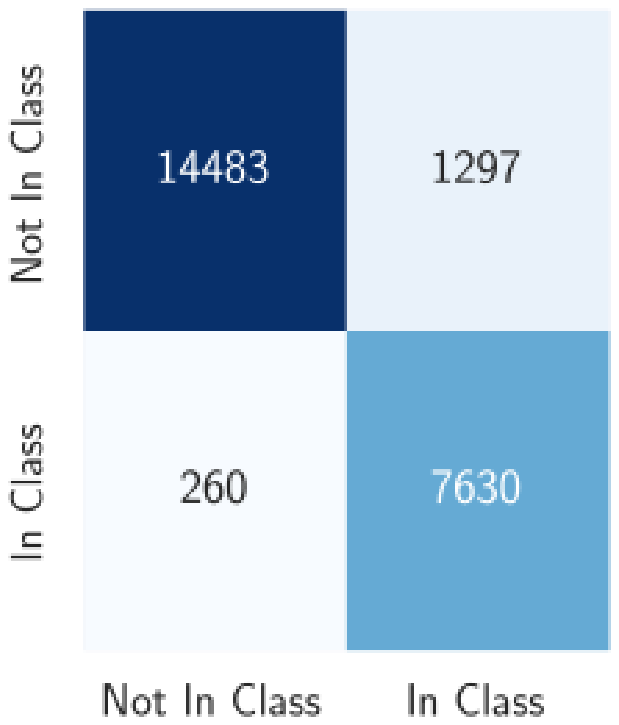} \hskip 0.1in
\includegraphics[height=1.5
in]{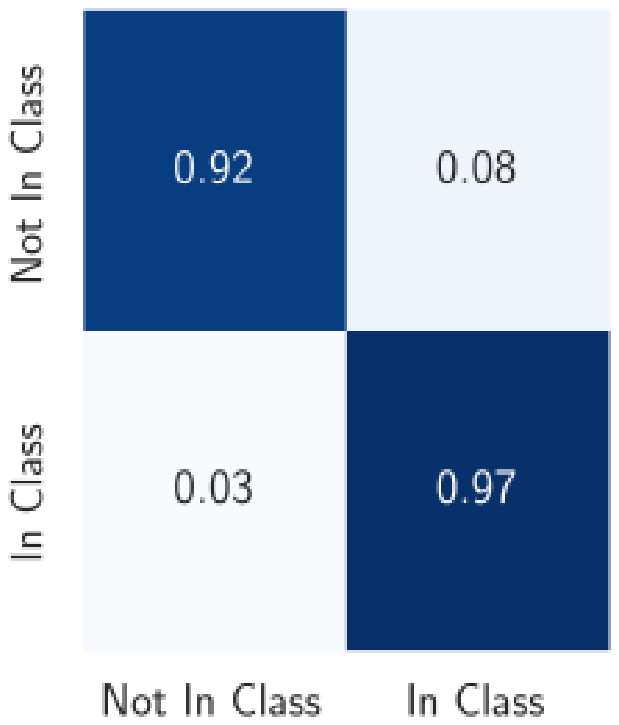}
\caption{\label{fig:worst-cm} Confusion matrix for the worst performing set of hyper-parameters for the SVC based on the macro-F1 score.. The matrix on the left gives the sum over all test runs of true negatives (top left), false positives (top right), false negatives (bottom left), and true positives (bottom right).  The matrix on the right shows the fractions for the raw values. In this case, the top row of the matrix shows the SVC correctly classified 92\% of points outside the class as being outside the class while incorrectly classifying 8\% of the points lying outside the class as belonging to the class. On the bottom row, the SVC incorrectly classified 3\% of the class points as lying outside the class while 97\% of points in the class were correctly classified as such.}
\end{figure}

Figures (\ref{fig:best-cm}) and (\ref{fig:worst-cm}) are confusion matrices for settings of hyper-parameters for the SVC that resulted in best and worst performance based on macro-F1 score.  The randomized cross validation search procedure chose 10 values for the SVC hyper-parameters, and then ran the SVC at each of the 10 settings over all 30 folds.  The confusion matrices were constructed by summing all of the metrics for the best and worst performing SVC. For all settings of hyper-parameters the SVC performed equally well when labeling class points as belonging to the proper class.  On the other hand, there was more variation in performance of the SVC incorrectly classifying out-of-class points as belonging to the class.  However even with the worst performing set of hyper-parameters, the SVC correctly classified 92\% of out-of-class points as not belonging to the class.  The high level of performance for even the worst-performing SVC can be attributed to the lack of noise in the synthetic data set. The closer a point is to the decision boundary, the more difficult it is for the SVC to properly classify it.  The states shown in the top panel of figure (\ref{fig:svc-fmc}) are in contiguous regions of the phase diagram; i.e., few if any points from any one of the states encroaches into a region dominated by another state.  In other words, there is little to no overlap or pockets of states within other states.

\begin{figure*}[hbt!]
\includegraphics[height=2.5in]{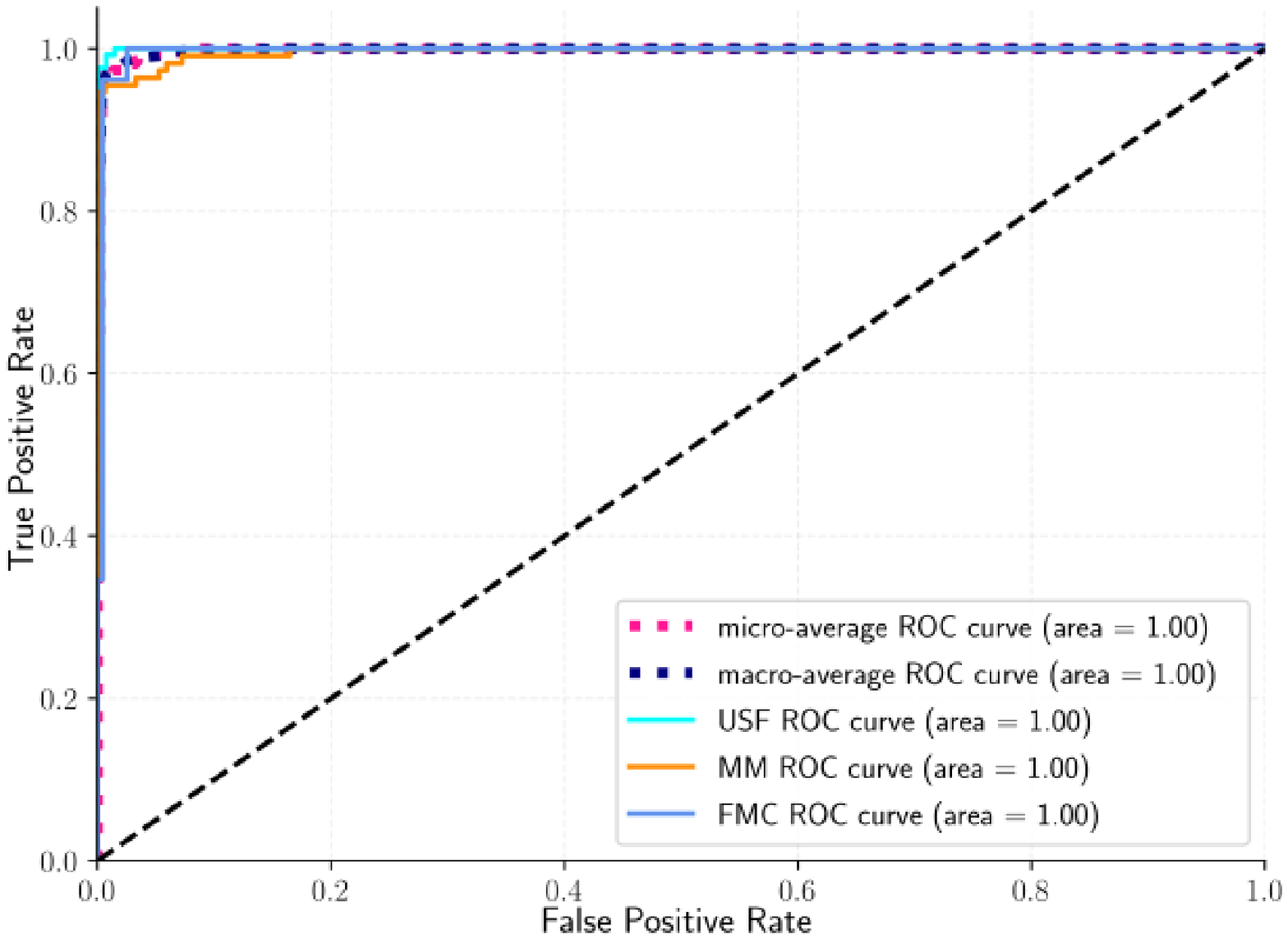} \hskip 0.1in
\includegraphics[height=2.5in]{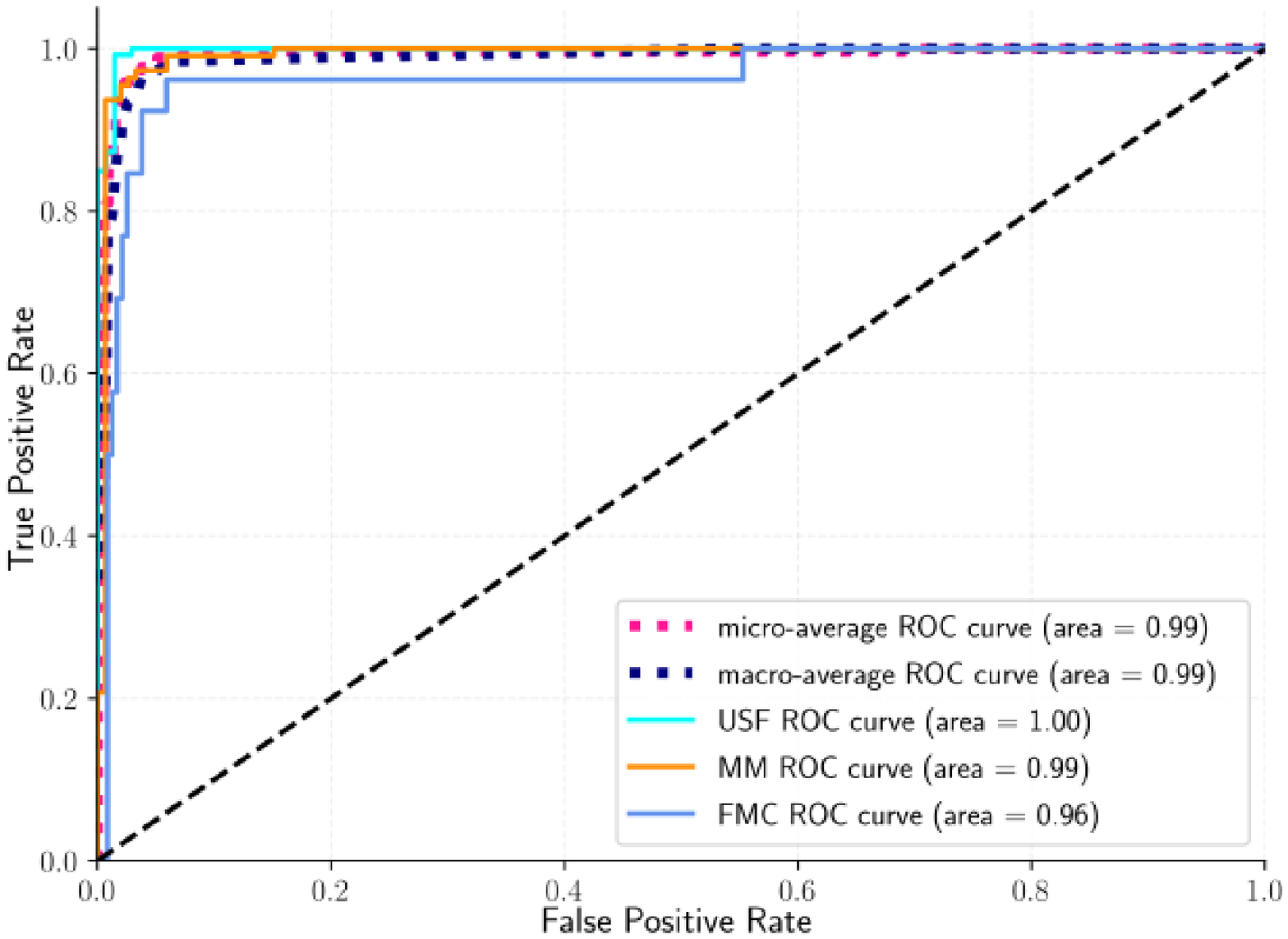}
\caption{\label{fig:roc-auc} Receiver operating characteristics curves for the folds with the largest area under the curve (left) and smallest area under the curve (right). The pink and blue dotted lines are the micro and macro-averaged ROC curves.  The micro-average is found by counting the total true positives (class is correctly identified), false negatives (point inside class is incorrectly identified as lying outside the class), and false positives (point outside class is incorrectly identified as being part of the class).  The macro-average is found calculating the average for each class, and then finding the unweighted mean.  The black dashed line at $False Positive Rate = True Positive Rate$ represents random chance for a binary classifier.}
\end{figure*}

The hyper-parameters for the SVC resulting in the best performance was then retrained over the same 30 stratified folds so that the receiver operating characteristics curves could be calculated.  Figure (\ref{fig:roc-auc}) show the folds where the SVC had the largest and smallest areas under the receiver operating characteristics curves for each of the three classifiers.  For the worst performance, we can see that the FMC classifier performed worse than the MM classifier, while in both best and worst cases, the USF classifier performed equally well. This can again be attributed to the lack of noise in the synthetic data.  There is virtually no overlap between FMC and USF states; however at different particle densities, the FMC melts into the normal MM state at different values of the Zeeman field resulting in an FMC that appears to protrude somewhat into the MM state.

\section{Conclusions}
We were able to show that a support vector classifier could be used to actively learn phase boundaries of a many-body quantum system.  Even though the full data set was created for another analysis and no timing measurements were taken, we can be reasonably confident that, since the SVC used only 7\% of this data to create remarkably accurate phase boundaries, we would have saved a significant amount of computation time.  

Referring again to fig. (\ref{fig:alp}), we emphasize there must be some physical intuition responsible for constructing a model, and a method or procedure to analyze it. There must also be an initial review of the output of the procedure to verify the output is sensible and to be reasonably expected.  Only after this initial work has been completed can an SVC be given the definitions of good and bad examples, and used to automate the review of output.

\section{Acknowledgements}
We would like to thank Professor Carlotta Domeniconi of the GMU Computer Science Department for her many helpful discussions and guidance. 

\FloatBarrier




\end{document}